\documentstyle[prc,preprint,aps]{revtex} 
\tighten 
\begin{document} 
\draft 
\title{Neutron-Proton Correlations in an Exactly Solvable Model} 
\author{J. Engel$^a$, S. Pittel$^b$, M. Stoitsov$^c$, P. Vogel$^d$, and 
J. Dukelsky$^e$}
\address{$^a$Department of Physics and Astronomy, University of North
Carolina, Chapel Hill, NC 27599, \\ 
$^b$Bartol Research Institute, University of Delaware, Newark, DE 19716, 
\\ $^c$Institute of Nuclear
Research and Nuclear Energy, Bulgarian Academy of Sciences, Sofia-1784, \\
Bulgaria and Bartol Research Institute, University of Delaware, Newark, DE
19716, \\ 
$^d$Department of Physics, Caltech, 161-33, Pasadena, CA 91125,\\ 
$^e$Instituto de Estructura de la Materia, Consejo Superior de
Investigaciones Cientificas, 
\\ Serrano 123, 28006 Madrid, Spain}
\date{\today} 
\maketitle

\begin{abstract}

We examine isovector and isoscalar neutron-proton correlations in an exactly
solvable model based on the algebra $SO(8)$.  We look particularly closely at
Gamow-Teller strength and double beta decay, both to isolate the effects of
the two kinds of pairing and to test two approximation schemes:  the
renormalized neutron-proton QRPA (RQRPA) and generalized BCS theory.  When
isoscalar pairing correlations become strong enough a phase transition occurs
and the dependence of the Gamow-Teller $\beta^+$ strength on isospin changes
in a dramatic and unfamiliar way, actually increasing as neutrons are added
to an $N=Z$ core.  Renormalization eliminates the well-known instabilities
that plague the QRPA as the phase transition is approached, but only by
unnaturally suppressing the isoscalar correlations.  Generalized BCS theory,
on the other hand, reproduces the Gamow-Teller strength more accurately in
the isoscalar phase than in the usual isovector phase, even though its  
predictions for energies are equally good everywhere.  It also mixes $T=0$ 
and $T=1$
pairing, but only on the isoscalar side of the phase transition.

\end{abstract} 

\section{Introduction}

Pairing correlations are an important feature of nuclear structure \cite{BM}.
In heavy nuclei such correlations between neutrons and protons are usually
neglected on the grounds that the two fermi levels are far apart.  In nuclei
with $N \approx Z$, however, the fermi levels are close and neutron-proton
($np$) pairing correlations can be expected to play a significant role in
nuclear structure and decay (for a review of work on $np$ pairing theory see
Ref.\ \cite{Goodman}).  The importance of these proton-rich nuclei in
astrophysical nucleosynthesis makes it vital that the $np$ correlations are
well understood, and upcoming experiments with radioactive beams will soon
test our understanding.

Taking $np$ correlations seriously complicates the usual treatment of
pairing, which stresses the interaction of like particles in time-reversed
orbits, i.e.\ the formation of $pp$ and $nn$ pairs.  Generalizing this
picture raises at least two new issues.  First, the $pp$, $nn$, and $np$
isovector pairs must all be treated on an equal footing so that isospin
symmetry is respected as much as possible\cite{ELV}.  Second, the competition
between two kinds of $np$ pairing --- isovector and isoscalar ($T=1$ and
$T=0$) --- must be taken into account.  This issue apparently arises even in
nuclei with $N ~>~ Z$ where $np$ pairing is by most measures small.  For
example, the rate of two-neutrino double-beta decay within the $np$
Quasiparticle Random Phase Approximation (QRPA) is extremely sensitive to the
strength of isoscalar particle-particle (i.e.\ pairing) correlations, making
reliable calculations difficult.  When these correlations become strong
enough the method fails even to give finite answers.

The breakdown in the QRPA signals an impending phase transition.  Is it real
or an artifact of the assumption underlying the approximation that the ground
state contains no $np$ correlations?  What are the properties of the
``isoscalar phase", if it is real?  Is the Renormalized QRPA
(RQRPA)\cite{Suhonen,Faessler1}, in which solutions are more stable, a good
way to handle the breakdown/phase transition?  Is generalized BCS theory, a
scheme for treating $np$ pairing on a more equal footing with $nn$ and $pp$
pairing, able to quantify the interplay between the two phases?  We address
these questions here in a solvable model that incorporates both isovector and
isoscalar pairing, making it considerably richer than the more schematic
models (e.g.\ the Lipkin model\cite{Lipkin}) typically used for this kind of
study.

The structure of this paper is as follows:  In section 2 we describe the
model and its analytic solution for energies and Gamow-Teller beta-decay
matrix elements, stressing the existence of two limiting solutions
corresponding to pure isovector or isoscalar pairing.  We show that in the
isoscalar phase, charge-changing processes have counterintuitive features.
Section 3 contains an outline of the QRPA and RQRPA as realized in the
solvable model, and applies them to single and double beta-decay to test the
quality of the approximations.  In Section 4 we describe generalized BCS
theory for $SO(8)$, again with emphasisis on beta decay strengths (though we
also examine ground-state energies), and again test the reliability of the
approximation scheme.  Section 5 is a conclusion.

\section{The Model and its Exact Solution}

We consider a set of degenerate single particle orbitals, characterized by
$l, s=1/2, t=1/2$.  The total number of single-particle states is $\Omega =
\sum_l (2l+1)$.  We make the model solvable by building a basis entirely from
$L=0$ operators:  $S_{\nu}^{\dagger}$, which creates pairs with spin $S=0$
and isospin $T=1$ (with projection $\nu$), and $P_{\mu}^{\dagger}$, which
creates a pairs with $S=1$ and $T=0$ (spin projection $\mu$).  Together with
the one body operators that generate $SU(4)$ --- the total spin $\vec{\cal
S}$, the isospin $\vec{\cal T}$, and the operator ${\cal F}^{\mu}_{\nu} =
\sum_i \sigma(i)_{\mu} \tau(i)_{\nu}~$ --- the $L=0$ pair creation and
annihilation operators form the algebra $SO(8)$\cite{Pang}.  The physics
associated with this model has been studied
previously\cite{Pang,Dussell1,Dussell2}, but with emphasis on energy levels;
here our focus will include charge-changing decay.

The most general Hamiltonian invariant under $SO(8)$, omitting terms such as
$\vec{S} \cdot \vec{S}$ and $\vec{T} \cdot \vec{T}$ that affect energies but
not wave functions, depends on three parameters and has the 
form\cite{Dussell1}
\begin{equation}
H = -\frac{g (1+x)}{2} \sum_{\nu} S_{\nu}^{\dagger} S_{\nu} - 
\frac{g(1-x)}{2}
\sum_{\mu} P_{\mu}^{\dagger} P_{\mu} +g_{ph} {\cal 
F}^{\mu\dagger}_{\nu}~{\cal 
F}^{\mu}_{\nu}~.
\end{equation}
The first term in the Hamiltonian corresponds to isovector spin-0 pairing, 
the second represents isoscalar spin-1 pairing, and the last is a (primarily)
particle-hole force in the $T=1$ $S=1$ channel. 

In certain important limits, analytic expressions for energies and wave
functions have been derived.  If $x=1$ and $g_{ph} = 0$, the
Hamiltonian\cite{Pang} conserves an $SO(5)$ subalgebra and corresponds to
``standard" spin-singlet isovector pairing, with $np$ pairs treated on an
equal footing with like-particle pairs\cite{Hecht1,Ginocchio}.  The
eigenstates, characterized by the number of nucleon pairs ${\cal N}$ (we
consider only nuclei with an even number of nucleons), the isospin $T$, and
the singlet-pairing seniority $v_s$, have energies
\begin{equation}
E(v_s,T) = -\frac{g}{8}\left[ (2{\cal N}-v_s)(4\Omega + 6 - 2{\cal N} - v_s)-
4 T(T + 1) \right]~.
\end{equation}
Similarly, for $x=-1$ and $g_{ph} = 0$ the exact solutions are characterized
by the spin $S$ and the triplet-pairing seniority $v_t$, and an analogous
formula applies with $T \rightarrow S, v_s \rightarrow v_t$.  This is the
``isoscalar phase" that will cause the breakdown of the QRPA.  Finally, if
$x=0$ the Hamiltonian is invariant under $SU(4)$.  The eigenstates are then
labeled by a quantum number $\lambda$ corresponding to the irreducible
$SU(4)$ representation $[\lambda,\lambda,0]$ as well as by $S$ and $T$, and 
the eigenvalues are
\begin{equation}
E(\lambda,S,T) =
-\frac{g}{4} \left[2{\cal N} (\Omega+3) - {\cal N}^2 - \lambda (\lambda+4)
\right] + g_{ph}\left[\lambda(\lambda+4)-S(S+1) -T(T+1)\right] ~.
\end{equation}
$S+T$ must be even if ${\cal N}$ is even and odd otherwise (this is true no 
matter what the Hamiltonian). The quantum number $\lambda$ has values  
$\lambda = S+T, S+T + 2, ...,\lambda_{Max}$, where $\lambda_{Max} = {\cal N}$ 
if ${\cal N} \le \Omega$ and $2\Omega-{\cal N}$ otherwise.

The eigenvalues and eigenstates of the general hamiltonian in Eq.\ (1) can be
obtained by diagonalizing in the $SU(4)$ basis.  The matrices are tridiagonal 
and have very small dimension.  Expressions for the matrix elements (with
several  typos) appear in Ref.\ \cite{Dussell1}, which, however, ignores the
particle-hole interaction.  The same model with the particle-hole interaction
included was solved approximately in Ref.\ \cite{Martin}.

By varying the parameter $x$ one can study the phase transition from the
standard spin-singlet isovector pairing phase, through the $SU(4)$ Wigner
supermultiplet phase, into the spin-triplet isoscalar pairing phase.  In
Fig.\ 1 we show the effects of this transition (with $g_{ph}$ fixed at zero)
on the overlap between the ground state and the ground state of the standard
spin-singlet paired system with $x=1$ and $g_{ph} = 0$.  [The abcissa is
labeled by $-x$, so that the more familiar isovector phase is on the left.]
The change produced by finite $g_{ph}$ is illustrated in Fig.\ 2; when
$g_{ph} > 0$ increases, the Hamiltonian more nearly conserves $SU(4)$
symmetry and the phase transition becomes less pronounced.  The overlap is an
obvious ``order parameter" in the model, and its point of inflection locates
the phase transition.  In even-even systems ($S=0$, $T$ even) this point
shifts to the right from the $SU(4)$-limit value $x=0$ as $T$ increases.  The
reason is that any excess neutrons are necessarily in isovector pairs, making
the transition to isoscalar pairing more difficult for the remaining
nucleons.

We turn our attention now to transitions induced by the GT operators ${\cal
F}^{\mu}_{\nu}$.  Their matrix elements are easily evaluated when the wave
function is written in the basis $|\cal{N},\lambda,S,T\rangle$ because the
operator is diagonal in $\cal{N}$ and $\lambda$ and can change $S$ and $T$ by
one unit only.  Explicit formulae can be constructed from the $SU(4)/SO(4)$
Clebsch-Gordan coefficients derived in Ref.\ \cite{Hecht}.  The GT operators
either increase or decrease $N-Z$ (assumed to be nonnegative); the
corresponding strengths are called $\beta^+$ and $\beta^-$.  The two
strengths are constrained by the Ikeda sum rule
\begin{equation} 
S(\beta^-) - S(\beta^+) = 3 (N-Z) ~. 
\end{equation} 
In the $SU(4)$ limit the $\beta^+$ strength vanishes and one state exhausts 
the $\beta^-$ strength.  In the two extreme limits surrounding $SU(4)$, i.e.\ 
$x=1$, $g_{ph} = 0$ (the isovector pairing phase) and $x=-1$, $g_{ph} = 0$ 
(the isoscalar phase) we can derive analytic expressions for $S(\beta^+)$ 
from the ground state as a function of $T=T_z = (N-Z)/2$, since in those 
limits the Hamiltonian contains only the generators of an $SO(5)$ subgroup.  
The $\beta$-decay operators break one pair (of either kind), leading to 
matrix elements between simple $SO(5)$ representations, the properties of 
which were studied in Ref.\ \cite{Hecht2}.  In the spin-singlet isovector 
phase we find
\begin{equation} 
\label{e:s1}
S(\beta^+) = \frac{({\cal N}-T)(T+1)( 2 \Omega-{\cal N}-T)}
{(2T/3+1)(\Omega+1/2)}~. 
\end{equation} 
This result applies in a single $j$-shell with degeneracy $2 \Omega$ as well
as in the model discussed here, which necessarily contains at least 2
degenerate levels in the $j-j$ scheme ($j=l \pm 1/2$).  The $\beta^+$ 
strength is plotted as a function of $T$ for $\Omega=12$, ${\cal N}=10$ in 
Fig.\ 3.  Except for the initial plateau at low $T$ the behavior of the curve 
is qualitatively similar to that obtained in BCS theory, where the 
neutron-proton interaction is ignored. The gradual decrease in strength with 
$T$ is caused by Pauli blocking.

In the isoscalar phase the result is
\begin{equation}
\label{e:s2}
S(\beta^+) = \frac{({\cal N}-T)(T+1)(2 \Omega -{\cal N}-T)}{\Omega-T+1/2}~.
\end{equation}
Here the behavior of the strength as $T$ increases from 0 is surprisingly
different (see Fig.\ 3).  The substantial rise at first seems 
counterintuitive since the neutron excess is increasing.  Blocking is not the 
only factor at work, however.  For lower $T$ the effect is overcome by the 
collective behavior of the boson-like $S$ pairs in the final state.  

In Fig.\ 4 we examine the behavior of the strength $S(\beta^+)$ between the
two limits, as a function of $-x$ for fixed $g_{ph} = 0$ (When $g_{ph} \ne 0$
all curves become flatter, because the system is closer to the $SU(4)$
limit.)  As $-x$ increases, the strength $S(\beta^+)$ decreases, vanishing
when the $SU(4)$ limit is reached and increasing again as the isoscalar
pairing phase is approached until finally it is considerably larger than in
the isovector phase.  The large strength is caused in part by the transfer of 
protons from $pp$ pairs, which cannot participate in $\beta^+$ decay, to 
isoscalar $np$ pairs, which can.  Only close to $x=-1$, however are this 
effect and the parabolic isospin dependence fully present; when $-x$ is small 
the strength can be small as well and the isospin dependence complicated, as 
is apparent from the crossings of curves in Fig.\ 4.  A large $\beta^+$ 
strength (compared, e.g., to the Ikeda sum rule) therefore reflects very 
strong isoscalar pairing.  If real, it would have important consequences for 
r-p process nucleosynthesis.

\section{Double-Beta Decay, the QRPA, and the RQRPA}

We have stressed $\beta^+$ strength because of its simplicity and sensitivity
to details of nuclear structure.  Now, however, we want to discuss
modifications of the QRPA, the most frequent and controversial application of
which is to double-beta decay.  Actually, in a model with as few states as
this one, the $\beta^+$ strength from the ``final" nucleus ($f$) essentially
determines the double-beta decay matrix element (considered here in the
closure approximation for simplicity and because the energy denominator can
change without a concommitant change in the wave functions).  The reason is
that the matrix element has the form
\begin{equation}
M_{GT}^{2\nu}(cl) =  \langle 0_f^+ | \sum_{i,j}\vec{\sigma}(i) \cdot 
\vec{\sigma}(j) \tau(i)_- \tau(j)_- | 0_i^+\rangle ~.
\end{equation}
and for a moderate neutron excess the $\beta^-$ strength, the other relevant
quantity, hardly varies with $x$.  In realistic calculations, the QRPA, which
has many desirable features, suffers an unfortunate instability when $g_{pp}
\equiv (1-x) g/2$ becomes too large that manifests itself through infinite
values for both the $\beta^+$ strength and the double-beta-decay rate.  A 
number of remedies have been proposed recently. One that has received 
particular attention is the Renormalized QRPA 
(RQRPA)\cite{Suhonen,Faessler1}, which eliminates the instability of the QRPA 
through a self-consistent calculation of the ground state.  The 
model presented here is ideal for examining how both QRPA and the new 
approximation work.

The p-n QRPA, described, for example, in Ref.\ \cite{Martin} and applied to
$SO(8)$ in the same paper, begins with the ordinary BCS ansatz, a coherent 
state of isovector neutron-neutron and proton-proton pairs, and proceeds by 
admixing neutron-proton quasiparticle pairs into the ground and excited 
states.  In $SO(8)$ the procedure leads to 2 by 2 matrix equations in each of 
the two (Fermi/isovector pairing and Gamow-Teller/isoscalar pairing) channels 
and can be solved by simple diagonalization.  More specificially, the one 
excited state in each channel ($S=0$ or $S=1$) is written in the form
\begin{equation}
\label{e:RPA}
|S> = \left( X_S [\alpha_p^\dagger \alpha_n^\dagger]^{L=0,S} - Y_S [\alpha_p 
\alpha_n]^{L=0,S} \right) |\tilde{0}> ~,
\end{equation}
where $\alpha_p^\dagger$, $\alpha_n^\dagger$ ($\alpha_p$, $\alpha_n$) create 
(destroy) proton and neutron quasiparticles, the brackets indicate angular 
momentum coupling, $|\tilde{0}>$ is the QRPA ground state 
and $(X_0$,$Y_0)$, and ($X_1$,$Y_1$) are the ``physical" eigenvectors in the 
spin-0, isospin-1 (Fermi) and spin-1 isospin-0 (Gamow-Teller) channels 
(additional details are in Ref.\ \cite{Martin}).  The two channels decouple 
and for two-neutrino double beta decay only the second is relevant.  
Associated with the $S=1$ eigenvector is an eigenvalue that becomes complex 
when $g_{pp}$ reaches a critical value connected with the impending phase 
transition.  The states with complex eigenvalues are not normalizable and 
have no physical significance, so that the approximation fails to give even 
an incorrect answer beyond the critical point.  This is the ``collapse" 
referred to above and is preceded by rapid changes in the $\beta^+$ and 
double-beta-decay amplitudes.

In the RQRPA, described for charge-changing modes in Ref.\
\cite{Suhonen,Faessler1}, the two channels are coupled in an attempt to make
the vacuum self consistent, and the resulting equations are nonlinear.  For
the $SO(8)$ model the equations have seven variables:  the two sets of $X$'s
and $Y$'s, the eigenvalue associated with each set, and a renormalization
parameter.  The iterative procedure advocated in Ref.\ \cite{Suhonen} often
does not converge here, but the model's simplicity makes the equations easy
to solve by other means.  Unlike the QRPA, the RQRPA never exhibits the
analog of the complex eigenvalues that signal instability, and therefore
never yields rapidly changing matrix elements.  The question is whether any
important physics is lost in the process of guaranteeing a ground state that
is built on the BCS state.

Fig.\ 5 presents the exact $\beta^-$ and $\beta^+$ strengths for fixed ${\cal
N}$ and $\Omega$ and several values of $T$, along with the QRPA and RQRPA
approximations to the strengths.  The QRPA breakdown is reflected in the
expression for the $\beta^+$ strength, which blows up at the critical value
of $x$ (or $g_{pp}$).  By contrast the RQRPA strength is perfectly stable.
The graphs make it clear, however, that the stability is achieved at a
significant price; the very real phase transition to a ground state dominated
by isoscalar pairing correlations changes the behavior of $S(\beta^+)$,
causing the QRPA to break down, but refuses to show itself at all in the
RQRPA approximation.  The reason is that in preserving (self-consistently)
the basic QRPA ansatz the RQRPA limits the isoscalar correlations in the
ground state.  Put another way, the QRPA breaks down for a reason; there
really is a phase transition and it really is nearby, and the RQRPA erases
all traces of it.  Thus at the very point at which the QRPA fails the RQRPA
also begins to deviate badly from the exact result.  To make matters worse,
and this has been noted elsewhere\cite{Hirsch}, renormalization destroys one
of the nicest features of the QRPA, the preservation of the Ikeda sum rule
Eq.\ (4). In this model, at least, nothing is gained by using the RQRPA.

To demonstrate this explicitly for double beta decay, we show in Fig.\ 6 the
matrix element $M_{GT}^{2\nu}$ for ${\cal N}=12$ and $T=4$ as a function of
$g_{pp}/g_{pair}$, where $g_{pair} = (1+x)/2$.  We use this parameter rather
than $x$ because it more closely resembles that used in realistic
calculations.  We have set $g_{ph} = 1.5 g$ so that the QRPA breaks down just
beyond the point at which the matrix element crosses the origin (when $g_{pp}
= g_{pair}$ i.e.\ at the $SU(4)$ point).  This is the situation in more
realistic calculations as well, but our model shows it to be pure
coincidence; the breakdown of the QRPA moves to larger $g_{pp}$ as the
essentially independent parameter $g_{ph}$ is increased, while the crossing
point never moves, implying that nothing fundamental is behind the proximity
of the crossing to the point at which the QRPA fails in realistic
calculations.  Interestingly, the exact matrix element in our model varies
smoothly as the phase transition is traversed, in sharp contrast to the rapid
drop in the prediction of the QRPA, which as noted blows up completely just
past the crossing point.  The RQRPA, as in the last figure, begins to fail at
the same point and offers little advantage over the QRPA itself.

In more realistic calculations this last conclusion may or may not hold.  The
simple model examined here contains only two very collective degrees of
freedom.  It is certainly possible that with less collectivity the QRPA
approximation is worse and the breakdown occurs far from the actual phase
transition.  In that event the RQRPA would offer advantages, especially in
the region between the breakdown of the QRPA and the real phase transition.
It would therefore be useful to examine the approximation in a model that
dilutes the collectivity of the $T=0$ $np$ pairs but is still solvable.  A
two-level version of $SO(8)$, i.e.\ $SO(8) \times SO(8)$\cite{Dussell2},
might be a good place to start.  Here we can say only that we find no
evidence supporting the validity of the RQRPA.

\section{Application of ``Generalized BCS Theory"}

In this section, we apply generalized pairing theory to the $SO(8)$ model, to
assess its ability to provide a meaningful approximate description of the
ground--state dynamics of the model in the various phases.  We simplify the
model slightly by setting $g_{ph}$ to zero (and $g$ to 2, which merely
scales the energies).

Generalized pairing theory is well reviewed in Ref.\ \cite{Goodman} and thus
will not be discussed in detail here.  Suffice it to say that the theory is
founded in the HFB approximation, supplemented by the further assumption that
the only nonzero matrix elements of the Hartree Fock and pair potentials are
those connecting the four states $|\alpha p>$, $|\alpha n>$, $|\bar{\alpha}
p>$ and $|\bar{\alpha} n>$, where $|\bar{\alpha}>$ denotes the state obtained
by time-reversal on the state $|\alpha>$.  As such, the theory naturally
accomodates all pairing modes on an equal footing.  This includes the usual
$p\bar{p}$ and $n\bar{n}$ pairing as well as $p\bar{n}$, $n\bar{p}$ and $pn$
pairing.  Note that here and in subsequent discussion we explicitly
distinguish the pairing of particles in ``the same orbit" (e.g., $np$) from
the pairing of particles in ``time--reversed orbits" (e.g., $n\bar{p}$).

We have chosen to formulate the theory in terms of the density matrix
$\rho(\alpha)$ and the pairing tensor $t(\alpha)$.  These matrices, after
invoking time-reversal invariance, take the parametrized forms
\begin{equation}
{\rho}(\alpha)=
\left( \begin{array}{cccc}
\rho_1               & \rho_0 e^{-i \theta}&  0    &  \rho_3 e^{-i \theta}  
\\
\rho_0 e^{i \theta} & \rho_2   & -\rho_3 e^{-i \theta}&  0                 \\
0                 &-\rho_3 e^{i \theta} & \rho_1     &  \rho_0 e^{i \theta} 
\\
\rho_3 e^{ i \theta} &  0       & \rho_0 e^{-i \theta} &  \rho_2         \\
\end{array} \right)_\alpha~, \mbox{~~~~~~~~}
 t(\alpha)=
\left( \begin{array}{cccc}
 0                & t_3 e^{ -i \theta} & t_1               & t_0 e^{-i 
\theta}\\
-t_3 e^{ -i \theta}& 0                 & t_0 e^{i \theta} & t_2               
\\
-t_1               &-t_0 e^{i \theta} & 0                 & t_3 e^{i \theta}  
\\
-t_0 e^{ -i \theta}&-t_2               &-t_3 e^{i \theta} & 0
\end{array} \right)_\alpha~,
\end{equation}
with the coefficients interrelated by four unitarity conditions

\begin{equation}
\begin{array}{c}
(1-\rho_1-\rho_2) \rho_0  - (t_1 + t_2) t_0 = 0~, \\
~\\
(1-\rho_1-\rho_2) \rho_3  + (t_1 + t_2) t_3 = 0~, \\
~\\
\rho_1 - \rho^2_1 -\rho^2_0 - \rho^2_3 - t^2_0 - t^2_1 - t^2_3 = 0~, \\
~\\
\rho_2 - \rho^2_2 -\rho^2_0 - \rho^2_3 - t^2_0 - t^2_2 - t^2_3 = 0~. \\
\end{array}
\label{unirt}
\end{equation}

In our application to the $SO(8)$ model, we impose constraints on the average
number of neutrons and the average number of protons of the system, thereby
fixing the parameters $\rho_1$ and $\rho_2$ according to
\begin{equation}
\displaystyle{
\rho_1= \frac{Z}{2\Omega} ~~~~~~\mbox{and}~~~~~~
\rho_2= \frac{N }{2\Omega} ~.}
\label{a12}
\end{equation}
Two more parameters, $\rho_0$ and $\rho_3$, are fixed from the first two of 
the unitarity conditions (\ref{unirt}).

Our procedure is first to express the energy of the generalized
quasiparticle vacuum as a function of the remaining five parameters of the
density matrix and pairing tensor and then to look for local minima, rather
than to solve the usual self--consistent eigenvalue equation.  The remaining
two unitarity conditions (\ref{unirt}) are implemented via Lagrange
multipliers.
The system of equations arising from these variational
conditions in principle admits several solutions (all having $\theta=0$ or
$\pi/2$), the energetically-lowest of which defines the generalized BCS
approximation to the ground state of the system.

Fig.~7 shows the energies associated with the solutions to the
generalized BCS equations for the case of $\Omega=12$ and ${\cal N}=5$.  The 
results are plotted as a function of the hamiltonian parameter $x$ and for 
various values of the neutron number $N$. The solutions displayed in the 
figure have the following character:
\begin{itemize}
\item
{\it Solution A} corresponds to pure $p\bar{p}$ and $n\bar{n}$ pairing.
\item
{\it Solution B} corresponds to pure $T=1$ $p\bar{n}$ and $n\bar{p}$ pairing. 
It only exists when N=Z, where it is precisely degenerate (but does not mix)
with solution $A$.  
\item   
{\it Solution C} in general involves both 
T=0 $p\bar{n}$ and $n\bar{p}$ pairing and T=1 $p\bar{p}$ and 
$n\bar{n}$ pairing.  The relative importance of these different pair 
correlations is dictated by the three parameters $t_0$, $t_1$ and $t_2$. To a 
good approximation, the first reflects
the number of collective T=0 $p\bar{n}$ and $n\bar{p}$ pairs, whereas the 
latter two reflect the number of collective $p\bar{p}$ and $n\bar{n}$ pairs,  
respectively. As $x$ increases from -1,  the solution eventually merges 
into solution $A$, ceasing to exist beyond that ``critical point".  
At precisely this point, there is a change in the character of the ground 
state predicted in generalized BCS approximation that 
mirrors the true state of affairs. 
\end{itemize}
                                 
Figure~8 shows the generalized pairing results for the ground state energy in
comparison with the exact energies discussed earlier, again for $\Omega=12$
and ${\cal N}=5$.  Included are results corresponding to values of the
neutron particle number $N$ ranging from $N=5$ to $N=10$.  The results for
$N=0$ through $N=4$ follow from the symmetry of the problem.  The 
generalized pairing results correctly reproduce the trends of the exact
results, equally well in both phases.  The good
predictions hold up for even--even nuclei, odd--odd nuclei, symmetric nuclei
with $N=Z$, and nuclei with $N=0$ or $Z=0$.  The generalized pairing
approximation even reproduces the gradual shift of the phase transition to
negative values of $x$ when the difference between N and Z increases.

We now turn one last time to Gamow-Teller matrix elements, focusing here
(as above) on the summed $\beta^+$ strength.  We obtain the strength by
evaluating the quasi--particle vacuum expectation value of the 1+2 body 
operator
\begin{eqnarray}
\hat{S}_{\beta^+} = &  3~\sum_{l_1,m_{l_1},m_{s_1},l_2,m_{l_2},m_{s_2},\mu} 
~ (-)^{\mu} ~
(1 ~\mu ~\frac{1}{2}~ m_{s_1} | \frac{1}{2} ~ m_{s_1} + \mu )
(1 ~-\mu ~\frac{1}{2}~ m_{s_2} | \frac{1}{2} ~ m_{s_2} - \mu ) \nonumber \\
&~~ p^{\dagger}_{l_2,m_{l_2},m_{s_2}-\mu} ~~ n_{l_2,m_{l_2},m_{s_2}}
~~ n^{\dagger}_{l_1,m_{l_1},m_{s_1}+\mu} ~~ p_{l_1,m_{l_1},m_{s_1}}~.
\end{eqnarray}
Here $p^{\dagger}~(p)$ creates (annihilates) a real proton and 
$n^{\dagger}~(n)$ 
creates (annihilates) a real neutron. When the ground state is dominated by 
ordinary $p\bar{p}$ and $n\bar{n}$ pairing and represented by solution $A$, 
the total $\beta^+$ strength is given by the usual formula
\begin{equation}
S_{\beta^+} =~  3Z ~-~ \frac{3NZ}{2\Omega} ~.
\end{equation}

When solution $C$ applies, the result is
\begin{equation}
S_{\beta^+} ~=~ 3Z ~-~\frac{3NZ}{2\Omega} ~+~ 4\Omega^2 \rho_0^2 
~-~2 \Omega (t_0)^2 ~.
\end{equation}
(The  $\beta^-$ strengths follow from the above and the Ikeda sum rule, which 
is preserved at the BCS level).
In Figure~9, we compare the exact (Eqs.\ (\ref{e:s1}) and (\ref{e:s2})) and
generalized BCS results for the summed $\beta^+$ strength as a function of
$T_z$ in the isovector and isoscalar limits when $\Omega=12$ and ${\cal 
N}=10$. Solution $A$, which is equivalent to the ordinary BCS product wave 
function, reproduces the general trends of the exact $\beta^+$ strength but 
differs significantly as $T_z \rightarrow 0$.  The results in the isoscalar 
limit are much better for all values of $T_z$, including $T_z=0$.  Only for 
energies are the BCS results equally good in both phases.

Figure~10 compares the exact and generalized BCS $\beta^+$ strengths for the
same $\Omega$ and ${\cal N}$ values as in Figure~9.  Here, however, we only
consider the results for $N=14$ and $Z=6$ ($T_z=4$), but for all values of
$x$.  Once again, the generalized BCS approximation provides a somewhat
better reproduction of the trends in the exact results when the system is
dominated by isoscalar pairing correlations.  The suppression of $\beta^+$
strength that shows up as the $SU(4)$ limit is approached, however, is not
described well by the generalized BCS formalism, either on the isoscalar or
isovector side.  More accurate results in this regime may require an
extension beyond our BCS approximation that fully accomodates the coexistence
of all the different pairing modes on both sides of the phase transition.

A striking feature of Fig.~10 is the rise of $\beta^+$ strength in the
isoscalar region as $-x$ increases to 1.  That this rise is correlated as
noted earlier with the transfer of nucleons from $pp$ and $nn$ pairs to $T=0$
$np$ pairs (in time reversed orbits --- we have dropped the overscore) is
apparent from Fig.~11 where we plot the number of collective pairs of
different types for the same system as in Fig.~10.  We show both the
generalized BCS results and the ``exact" results, using the standard
prescription for operators that roughly measure the number of collective
pairs\cite{ELV} (It is this prescription, which we believe can be improved by
a better treatment of Pauli effects, that is responsible for the appearance
of low-levels of $np$ pairs before the phase transition, where the
corresponding pairing tensor is zero.)  Part b) of the figure clearly shows
that in BCS theory the wave function does not change from its product form
until the phase transition is reached.  It also shows that the constant wave
function is actually not so bad an approximation when $T_z$ is as large as in
the figure.  Only when $N \approx Z$ does the theory, which will not
reflect $np$ pairing to the left of the phase transition, fail badly.  Even
then, however, the BCS approximation manages to reproduce energies well.

\section{Summary and Concluding Remarks}

We have examined the interplay between isovector and isoscalar pairing modes
in an exactly solvable $SO(8)$ model, focusing on the matrix elements of
charge-changing operators.  The behavior of the $\beta^+$ strength in the
isoscalar phase is counterintuitive, rising with increasing neutron excess
instead of falling.  Double-beta decay, by contrast, varies smoothly and
predictably on both sides of the phase transition.  This behavior is in sharp
contrast to the predictions of the QRPA, which fails when isoscalar pairing
becomes too strong.

Partly for this reason, we have tested two approximation schemes that purport
to better accomodate neutron-proton correlations.  One, the RQRPA, works well
in the isovector phase but fails completely to capture the physics of the
phase transition.  Ironically the second approximation, provided by
generalized BCS theory, does a better job for the total beta-decay strength
in the isoscalar phase than in the isovector phase.  (It successfully
reproduces ground state energies everywhere, however, even in the vicinity of
the transition).  The reason is that the strength operator is a scalar in
space and spin, and is therefore not sensitive to the spin ``deformation"
that inhabits the BCS wave function in the isoscalar phase.  On the other
hand, the isospin deformation in the isovector phase distorts the expectation
value of the strength operator, which contains isoscalar, isovector, and
isotensor pieces.  In Ref.\ \cite{ELV} it was shown in a simpler model, based
on $SO(5)$ and containing only isovector pairing interactions, that
projection of the generalized BCS quasiparticle vacuum onto states with good
isospin after variation can fix this problem (though there the analog of
solution $B$ was used as the ``isointrinsic state").  It is far from clear,
however, that projection will allow the dynamical mixing of isoscalar and
isovector pairing before the phase transition is reached.
Recently\cite{Wyss}, the Lipkin-Nogami method was shown to do the trick at
least in part, and it would be interesting to test it in a model like this
one.  On the other hand, the phases are mixed even at the BCS level on the
isoscalar side of the critical point.

The shortcomings of the RQRPA and the successes of generalized pairing theory
raise the following question:  Can the generalized quasiparticle vacuum be
used as a starting point for a ``generalized QRPA" that works even in the
region of the phase transition?  Something along these lines has been
attempted in Ref.\ \cite{Faessler}, but only after forcing the BCS to mix
isoscalar and isovector pairing in an artifical way.  Our BCS also appears
not to mix the two kinds of pairs except to the right of the critical point,
a region that is probably unphysical in nuclei that undergo double-beta
decay.  Perhaps a more fruitful approach, therefore, will be a more
self-consistent QRPA, in which the RPA and BCS equations are coupled.  Such a
procedure can rescue the Ikeda sum rule\cite{Dukelsky} and could conceivably
facilitate isoscalar-isovector mixing at the BCS level even to the left of
the critical point.  Other modifications of the basic BCS procedure,
including approximations to projection, may also mix the pairing modes
without complicating the method too severely.  Which of these approaches is
the simplest and most useful remains to be seen.

\acknowledgements

We would like to thank Professor B.R.\ Mottelson for encouragement.  This 
work
was supported in part by the National Science Foundation under Grant Nos.\
PHY-9303041, PHY-9600445  and INT-9224875, by the U.S.\ Department 
of Energy under grants
DE-FG05-94ER40827 and DE-FG03-88ER-40397, by NATO under Grant No.\ 
CRG.900466, by the Bulgarian National Foundation for Scientific Research 
under Contract
No.\ $\Phi$-527, by the DIGICYT (Spain) under contract no. PB95/0123,
and by the European Union under contract no. CI1$^*$-CT94-0072. 
One of the authors (MVS) would like to acknowledge the 
support of the Fulbright Foundation.

\begin{figure}
\vskip 0.2 cm
\caption{Overlaps between the ground state and the pure isovector
spin-singlet paired state vs.\ the parameter $x$ in the hamiltonian of 
Eq.\ (1), for $\Omega = 12, {\cal N}=10$ and $S=T=0$ (solid line), $S=0, T=2$ 
(long dashes), and $S=0, T=4$ (short dashes). Here and in Figs.\ 2, 
4, 5, and 10 the quantity $-x$ is used on the abscissa axis so that the 
standard isovector phase is on the left.}
\label{fig:overlap}
\end{figure}

\begin{figure}
\vskip 0.2 cm
\caption{ Same as Fig.\ 1, with $S=T=0$, except for different values 
of $g_{ph}$.  The solid line corresponds to  $g_{ph} = 0$, the long dashes 
to $g_{ph} = 1.0 g$, and the short dashes to $g_{ph} = 2.0 g$.}
\label{fig:overlapph}
\end{figure}

\begin{figure}
\vskip 0.2 cm
\caption{ The Gamow-Teller $\beta^+$ strength B(GT) vs.\ the initial neutron
excess $T_z = T$ for $\Omega = 12, {\cal N} = 10$.  The solid curve is for 
the pure isoscalar spin-triplet pairing phase and the dashed curve for 
the standard isosvector spin-singlet phase.}
\label{fig:test}
\end{figure}

\begin{figure}
\vskip 0.2 cm
\caption{The Gamow-Teller $\beta^+$ strength B(GT) vs.\ the hamiltonian
parameter $-x$, for $\Omega = 12, {\cal N}=10$, $S=0$ and 
$g_{ph} = 0$. The solid curve corresponds to $T=0$, the long dashes to $T=2$, 
and the short dashes to $T=4$.}
\label{fig:beta+}
\end{figure}

\begin{figure}
\vskip 0.2 cm
\caption{ The Gamow-Teller strength B(GT) for $\beta^-$ (upper parts of each 
panel) and
$\beta^+$ (lower parts of each panel) vs.\ $-x$.  The exact results are 
denoted by
the solid lines and the RQRPA results by the dotted lines.  Calculated for 
$\Omega = 12, {\cal N}=10$, $S=0$, and $g_{ph} = g$ and several values of the 
isospin $T$ labelling the corresponding panels.}
\label{fig:betajon}
\end{figure}

\begin{figure}
\vskip 0.2 cm
\caption{Double beta decay matrix element $M_{GT}^{2 \nu}$ vs. 
$g_{pp}/g_{pair}$. The exact results are denoted by the solid line, the RQRPA 
results by the dashed line, and the QRPA results by the dotted line. 
Calculated for $\Omega = 12$, ${\cal N}=12$, $S=0$, $T = 4$ and $g_{ph} = 1.5 
g$.}
\label{fig:mgtjon} 
\end{figure}

\begin{figure}
\vskip 0.2 cm
\caption{Energies associated with the different variational solutions to the
generalized BCS equations, for $\Omega = 12, {\cal N} = 5$.  The panels are
labelled by the neutron number $N$.  Here and in Fig.  8 the
abscissa is the hamiltonian parameter $x$ running from 1 to -1.  The standard
isovector paired state is on the left as in the other figures.}
\label{fig:BCSresults}
\end{figure}

\begin{figure} 
\vskip 0.2 cm
\caption{Comparison of exact ground state energy (solid curve)
with the approximate result obtained in the generalized BCS approximation 
(dashed curve).  The results are shown for $\Omega=12$ and ${\cal N} =5$, and 
for different values of the neutron number $N$.}
\label{fig:gsenergies}
\end{figure}

\begin{figure}
\vskip 0.2cm
\caption{The Gamow-Teller $\beta^+$ strength B(GT) for $\Omega=12$ and ${\cal 
N}=10$ as a function of the isospin $T$ in the two limiting phases.  Solid 
lines represent the exact solutions and dashed lines the generalized BCS 
solutions in the isovector (bottom two lines) and isoscalar (top two lines) 
phases.}
\end{figure}

\begin{figure}
\vskip 0.2cm
\caption{Exact (solid) and generalized BCS (dashed) Gamow-Teller $\beta^+$
strength B(GT) vs.\ the hamiltonian parameter $-x$ for $\Omega=12$, ${\cal N} 
=10$ and $T_z = 4$.}
\end{figure}

\begin{figure}
\vskip 0.2cm
\caption{``Numbers" of different types of pairs in the exact (a) and BCS (b) 
solutions as a function of $-x$, for $\Omega=12$, ${\cal N}=10$ and $T_z=4$. 
The solid line measures the number of $nn$ pairs, the dashed line the number 
of $pp$ pairs, the dot-dashed line the number of $T=0$ $np$ pairs, and the 
dotted line the number (nonzero only because of the rough definition of 
``number") of $T=1$ $np$ pairs.}
\end{figure}
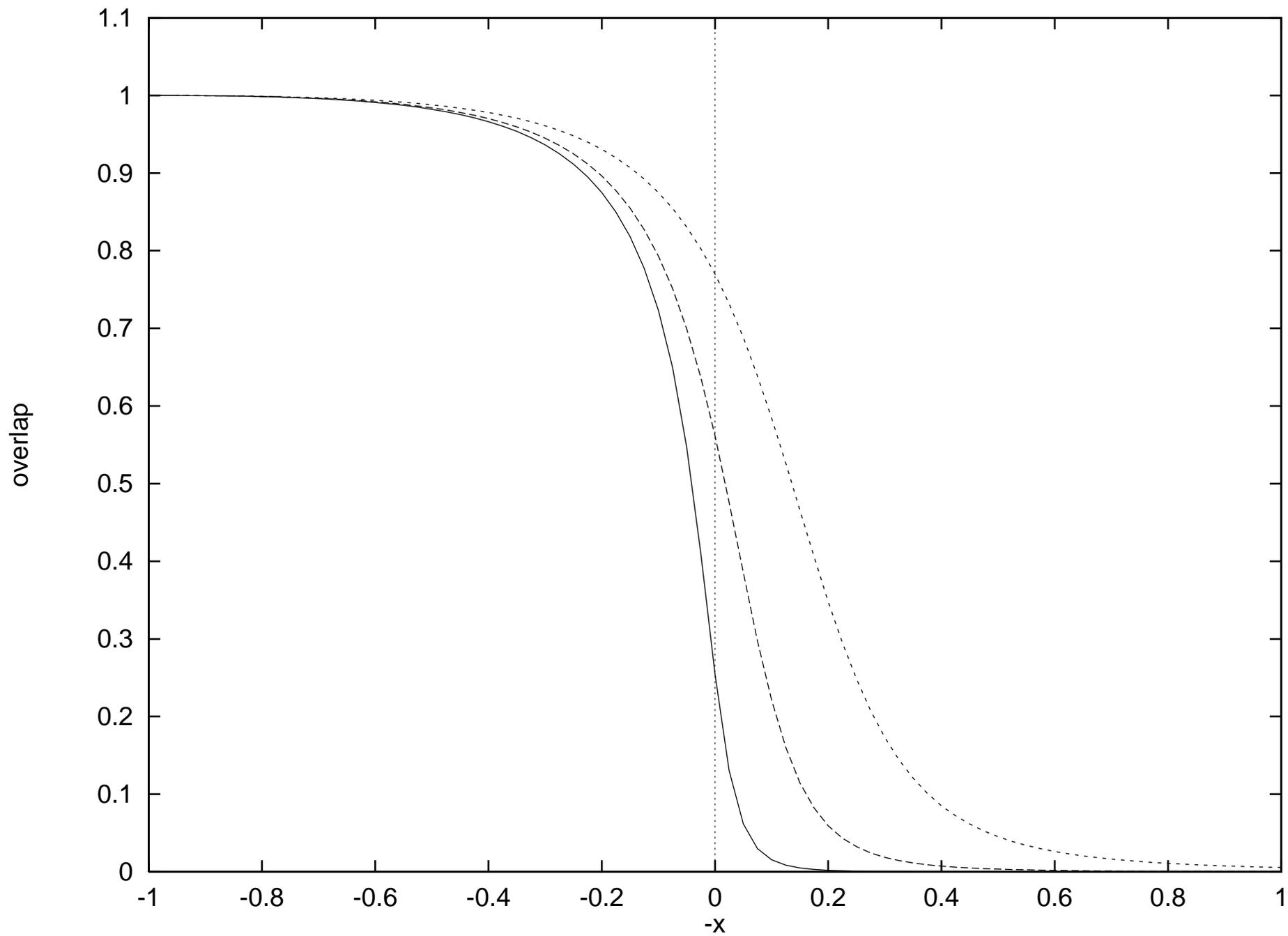
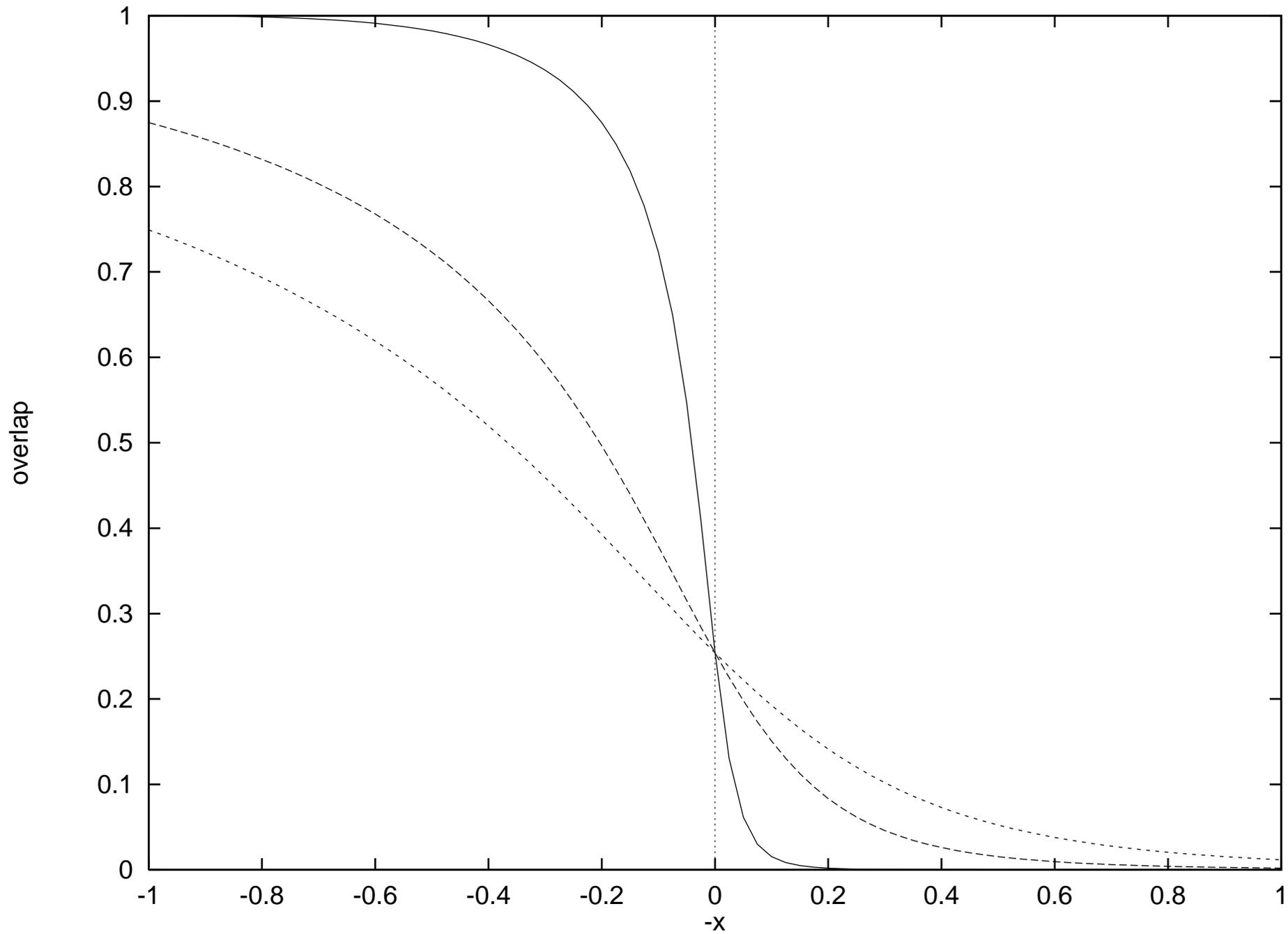
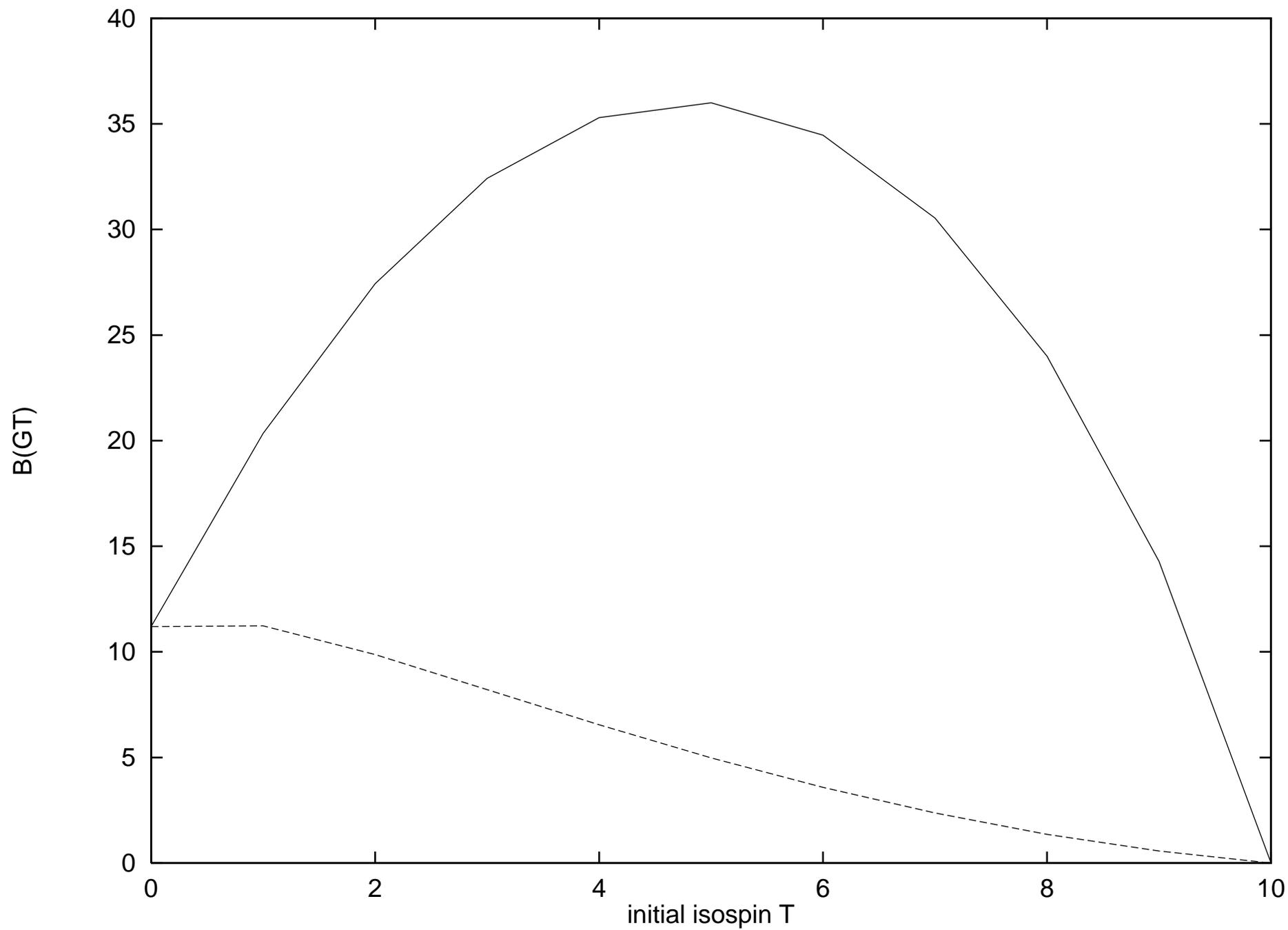
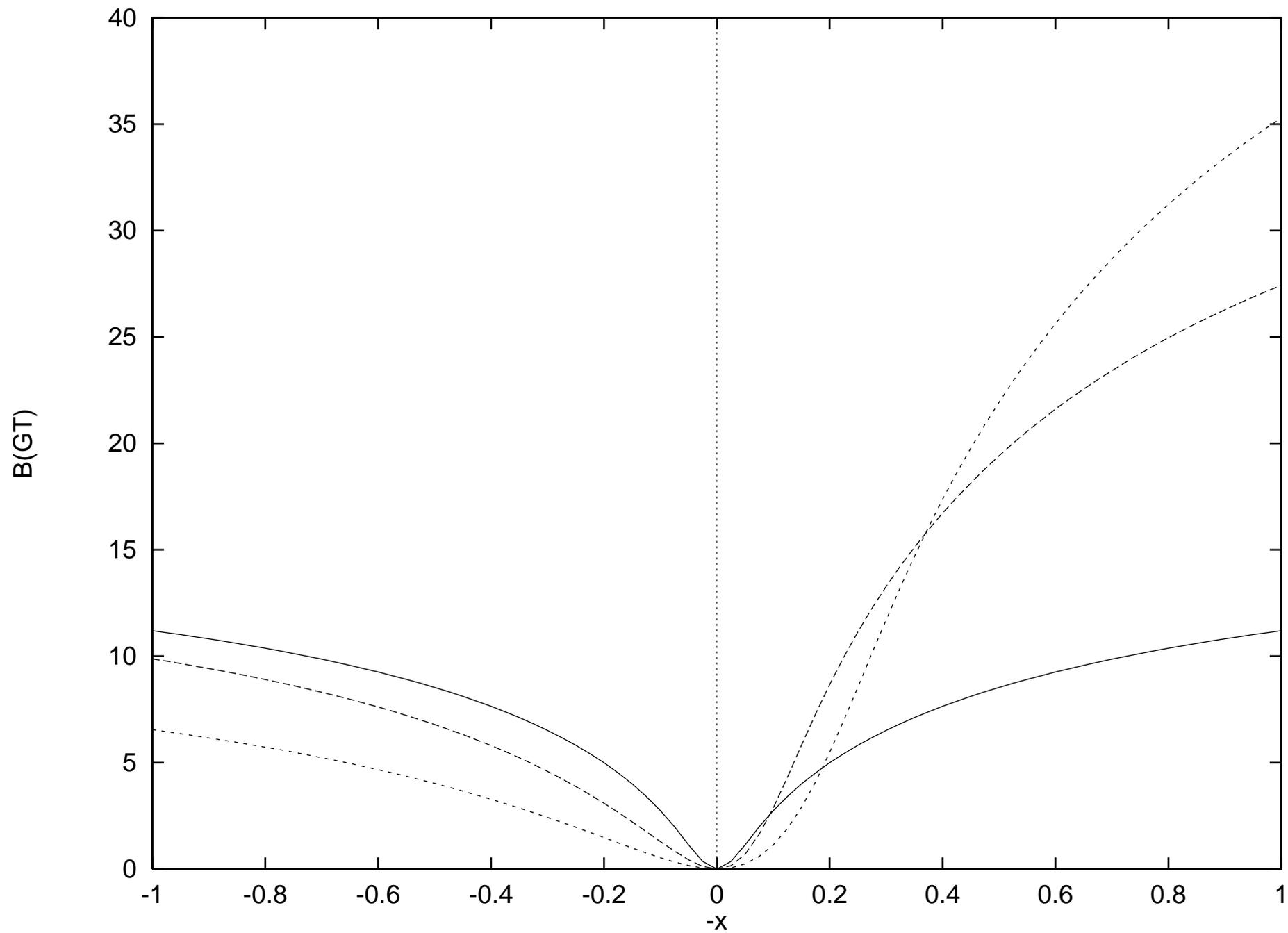
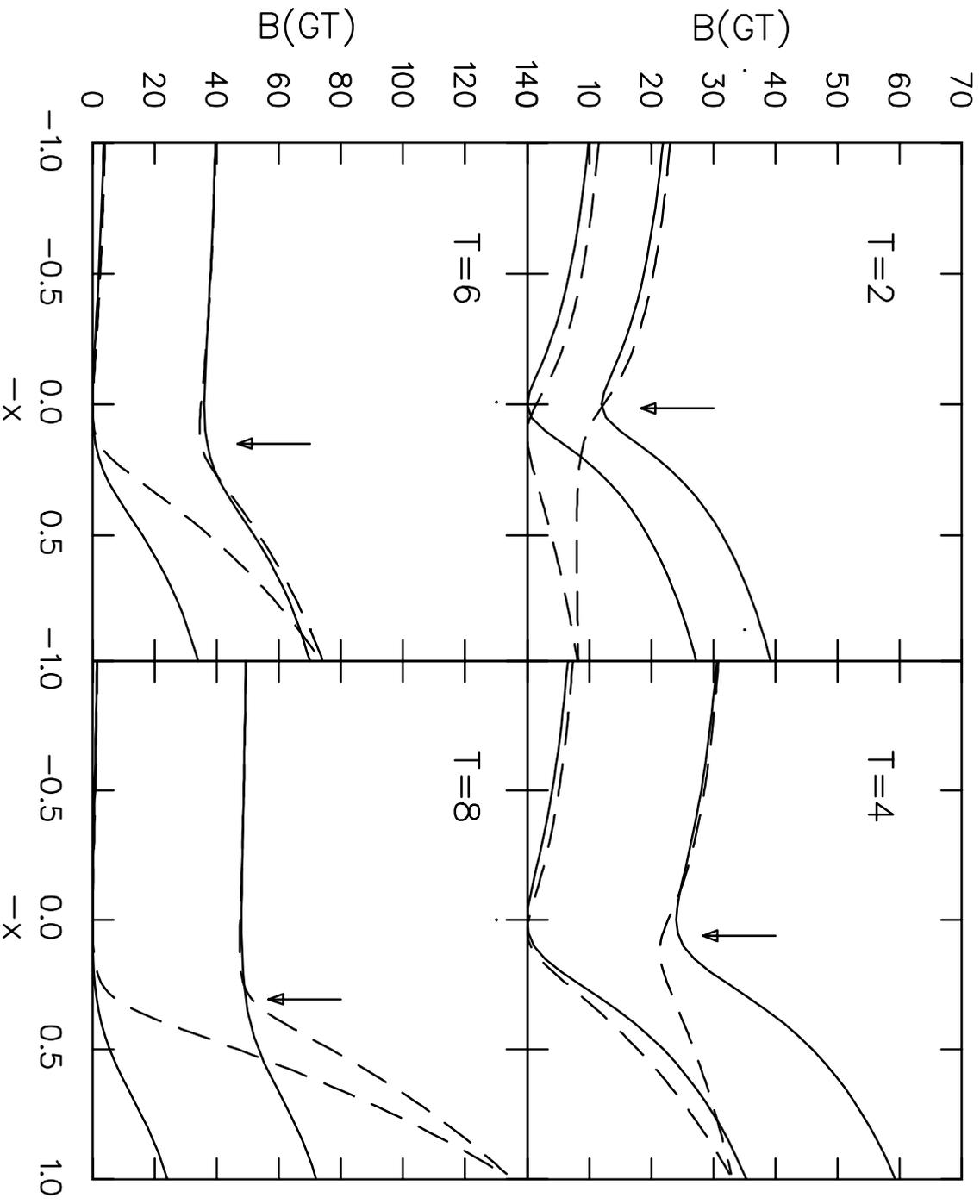
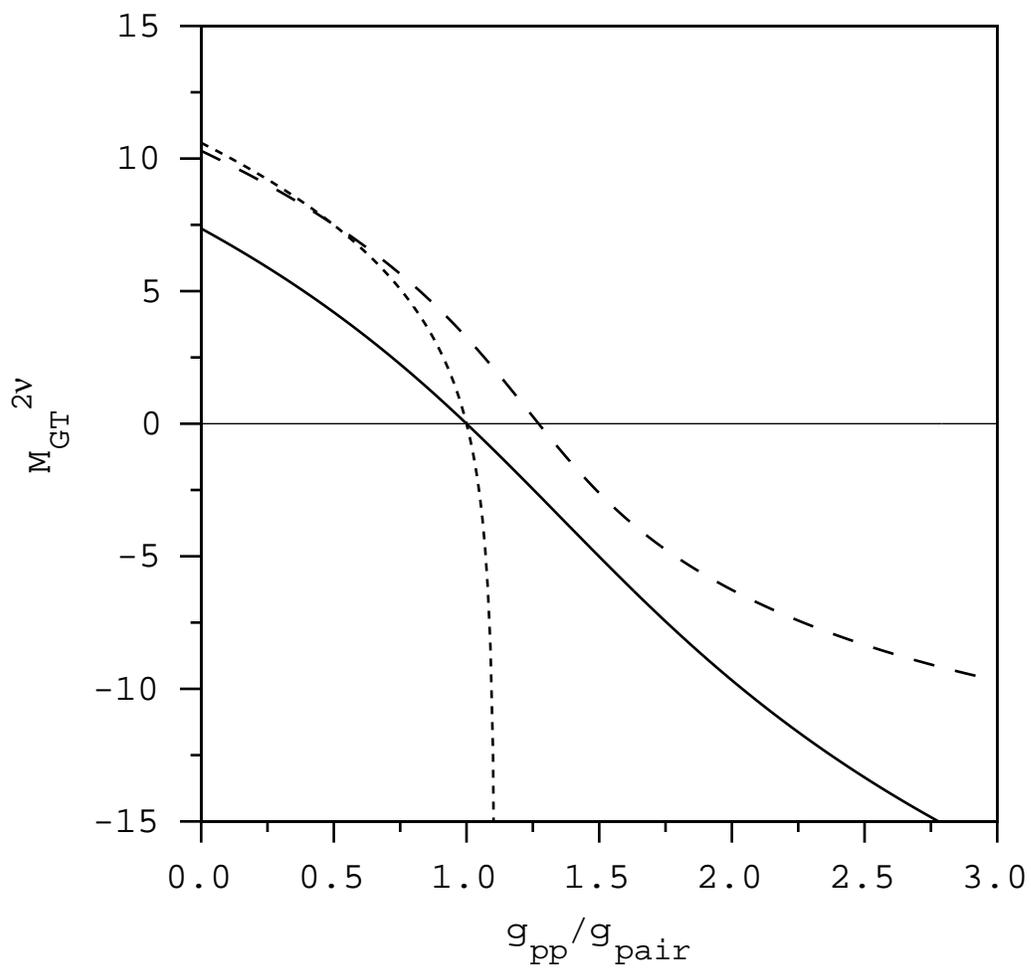
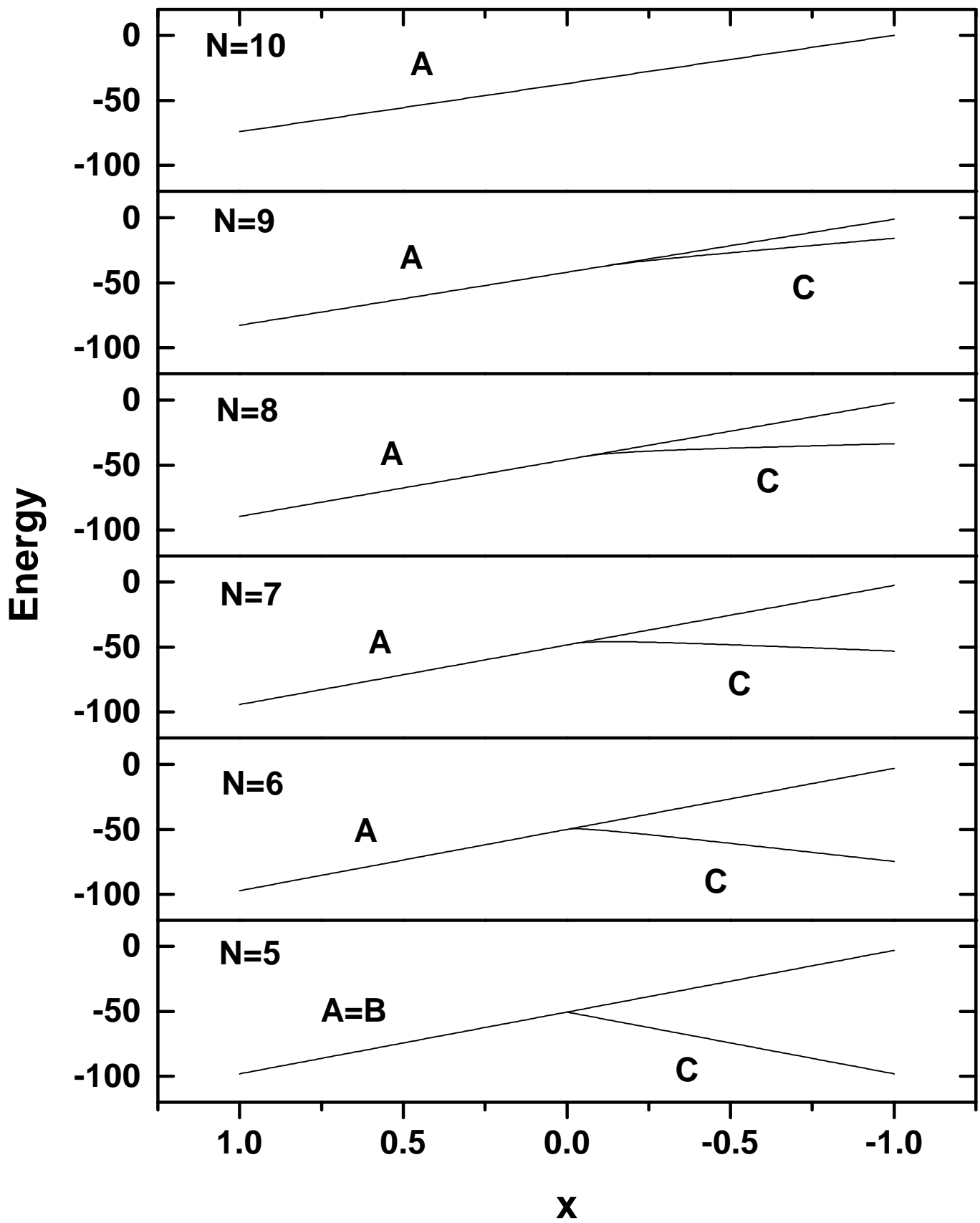
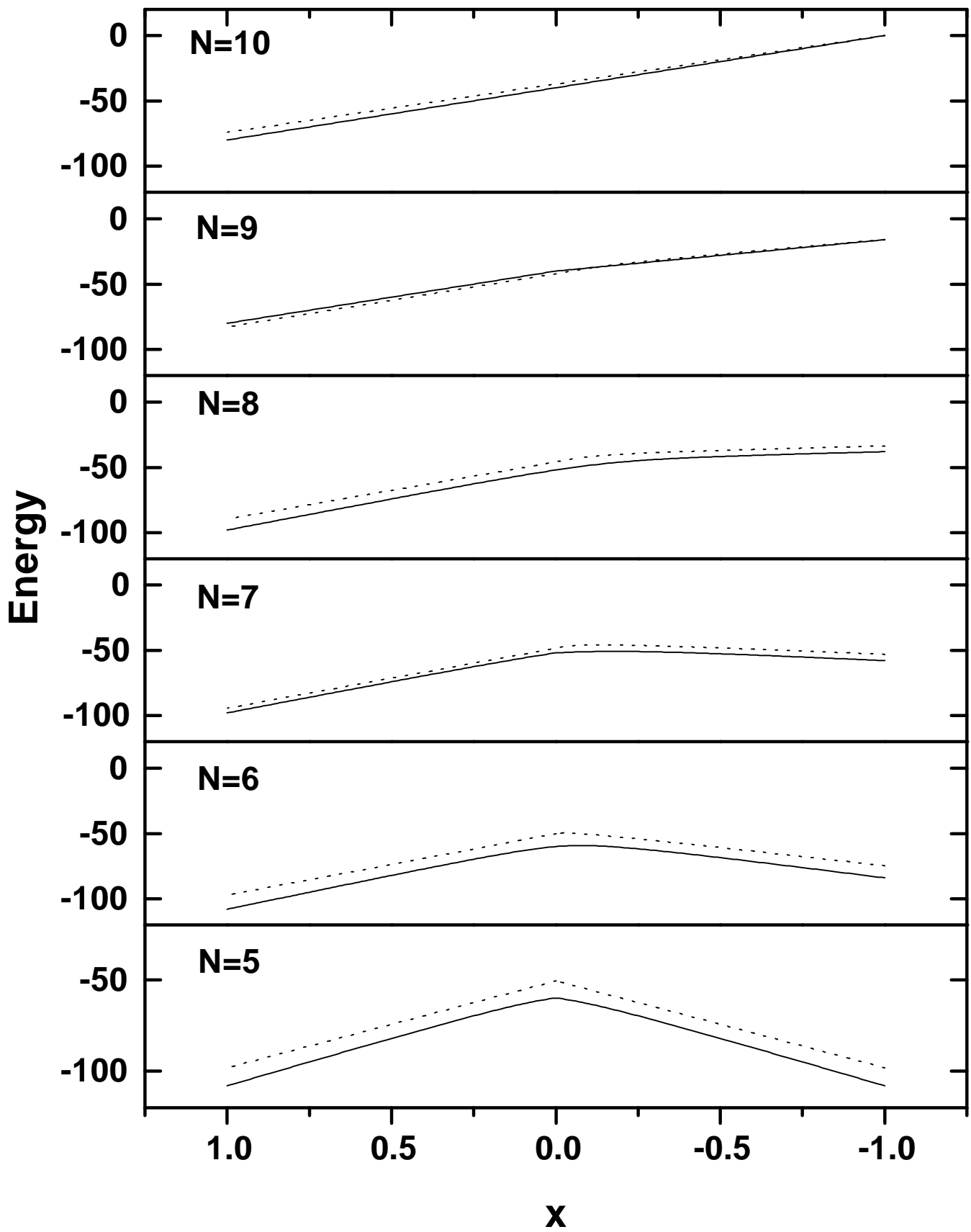
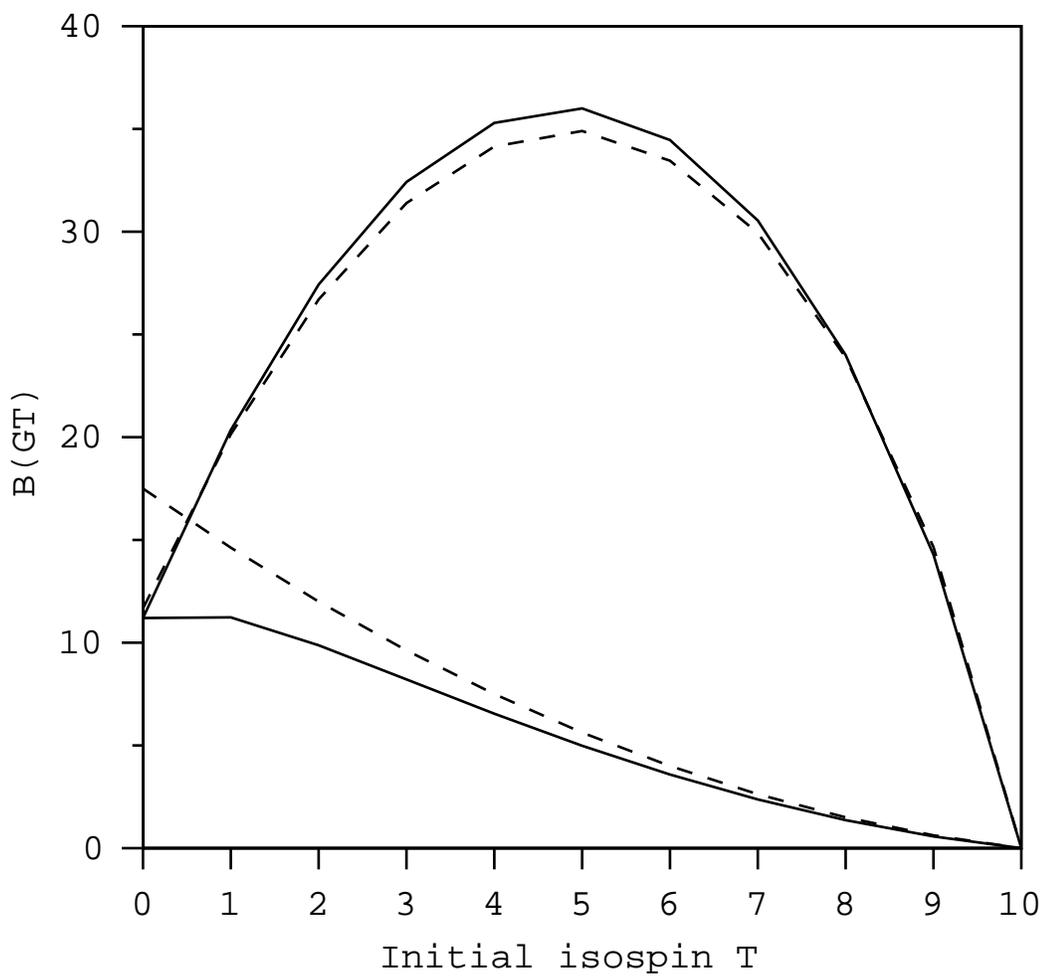
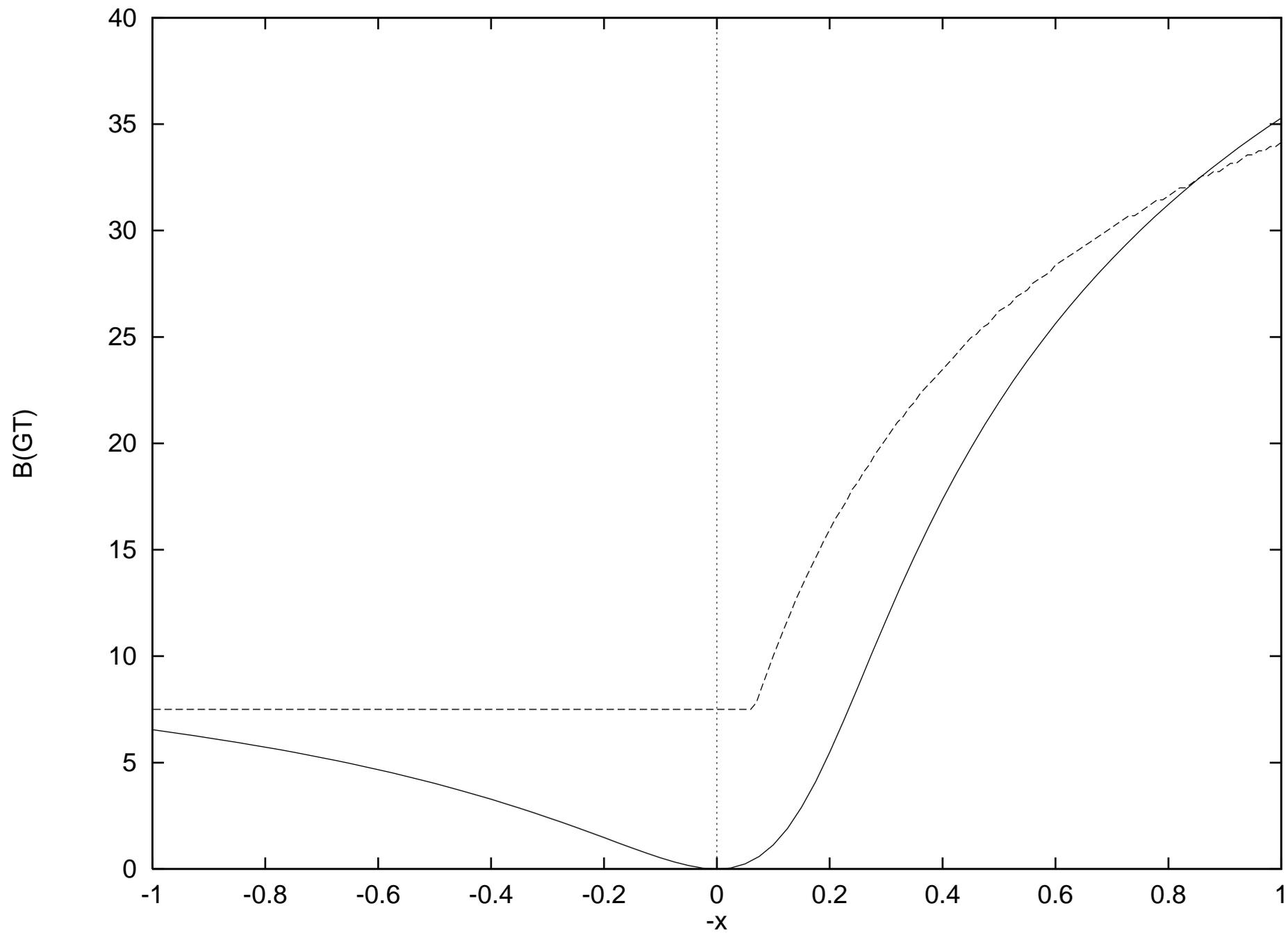
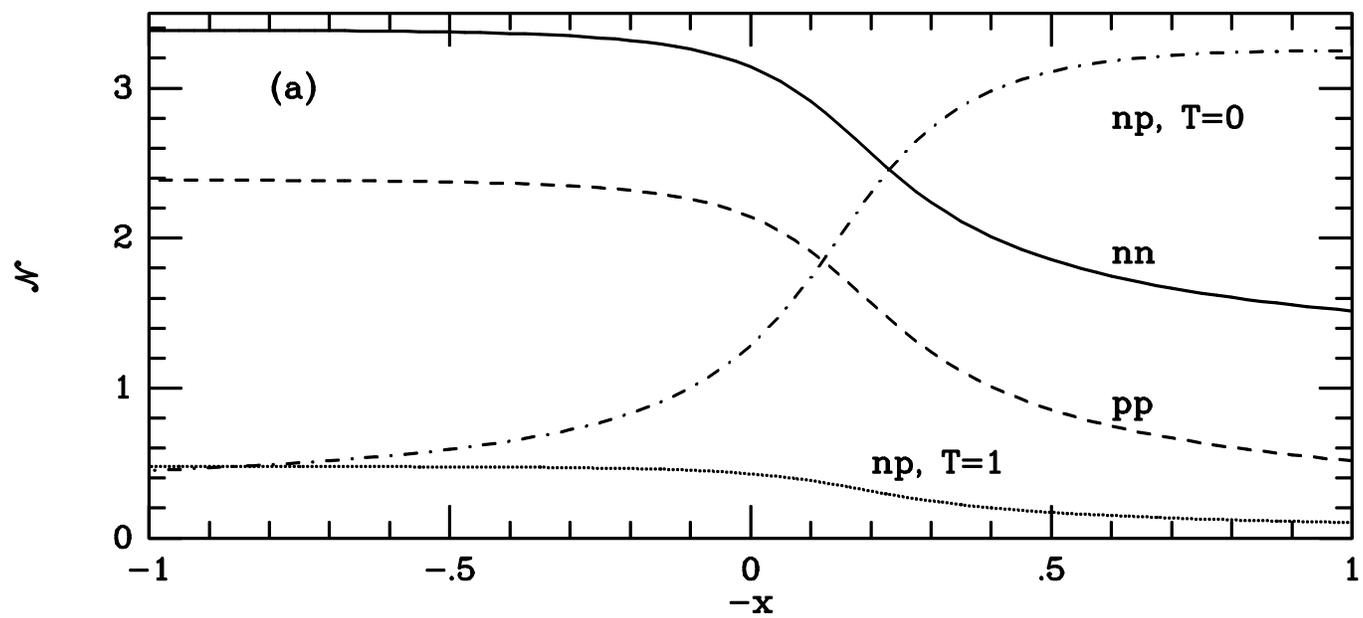
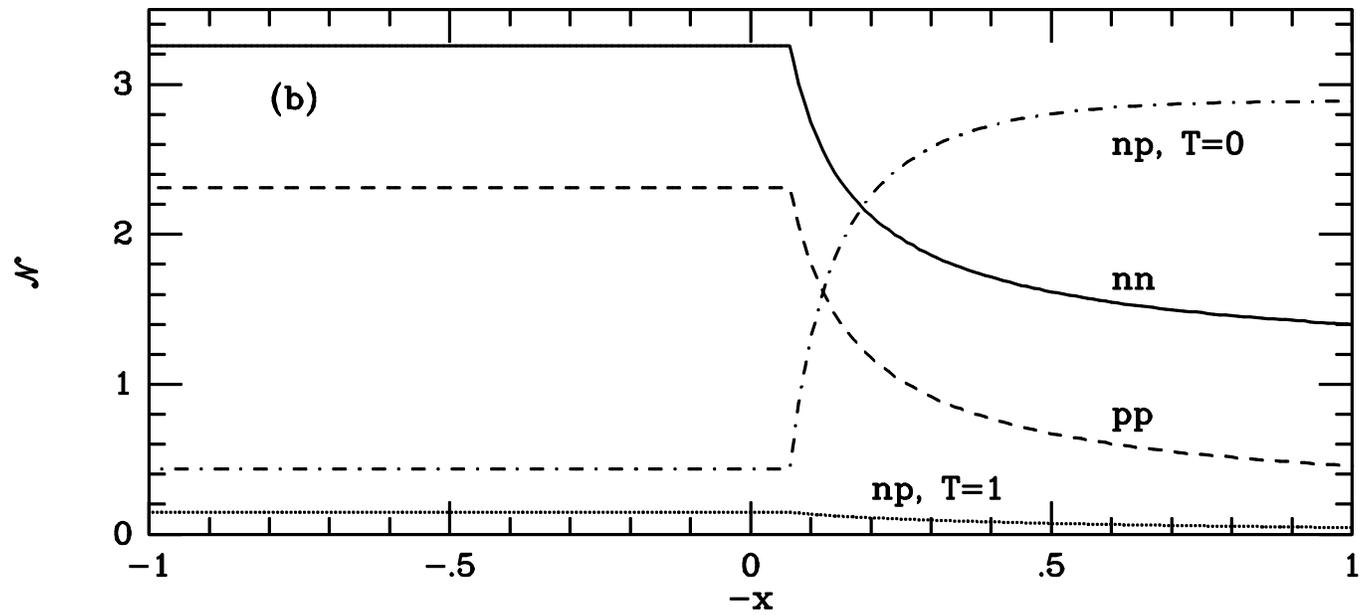
\end{document}